\documentclass[aps,prb,twocolumn,superscriptaddress]{revtex4-2}
\usepackage{graphicx}
\usepackage{amsmath}
\usepackage{amssymb}
\usepackage{mathtools}
\usepackage{xcolor}
\usepackage{braket}
\usepackage[colorlinks=true, citecolor=blue, urlcolor=blue, linkcolor=blue,bookmarks=false,hypertexnames=true]{hyperref} 
\usepackage{appendix}
\usepackage{array}
\usepackage[column=O]{cellspace}
\newcolumntype{P}[1]{>{\centering\arraybackslash}p{#1}}
\begin{document}
\graphicspath{}
\preprint{APS/123-QED}

\title{Electrically Tunable Fine Structure of Negatively Charged Excitons in Gated Bilayer Graphene Quantum Dots}
\date{December 11, 2023}

\author{Katarzyna Sadecka*}
\affiliation{Department of Physics, University of Ottawa,
Ottawa, K1N6N5, Canada}
\affiliation{Institute of Theoretical Physics, Wroc\l aw University of Science and Technology, Wybrze\.ze Wyspia\'nskiego 27, 50-370 Wroc\l aw, Poland}

\author{Yasser Saleem*}
\affiliation{Department of Physics, University of Ottawa,
Ottawa, K1N6N5, Canada}
\affiliation{Institut f\"{u}r Physikalische Chemie, Universit\"{a}t Hamburg, Grindelallee 117, D-20146 Hamburg, Germany}

\author{Daniel Miravet}
\affiliation{Department of Physics, University of Ottawa,
Ottawa, K1N6N5, Canada}

\author{Matthew Albert}
\affiliation{Department of Physics, University of Ottawa,
Ottawa, K1N6N5, Canada}

\author{Marek Korkusinski}
\affiliation{Department of Physics, University of Ottawa,
Ottawa, K1N6N5, Canada}
\affiliation{Security and Disruptive Technologies,
National Research Council, Ottawa, K1A0R6, Canada }

\author{Gabriel Bester}
\affiliation{Institut f\"{u}r Physikalische Chemie, Universit\"{a}t Hamburg, Grindelallee 117, D-20146 Hamburg, Germany}

\author{Pawel Hawrylak}
\affiliation{Department of Physics, University of Ottawa,
Ottawa, K1N6N5, Canada}


\begin{abstract}
We predict here the fine structure of an electrically tunable negatively charged exciton (trion) composed of two electrons and a hole confined in a gated bilayer graphene quantum dot (QD). We start with an atomistic approach, allowing us to compute confined electron and confined hole QD states for a structure containing over one million atoms. Using atomistic wavefunctions we compute Coulomb matrix elements and self-energies. In the next step, by solving the Bethe-Salpeter-like equation for trions, we describe a negatively charged exciton, built as a strongly interacting interlayer complex of two electrons in the conduction band and one hole in the valence band. Unlike in conventional semiconducting QDs, we show that the trion contains a fine structure composed of ten states arising from the valley and spin degrees of freedom. Finally, we obtain absorption into and emission from the trion states. We predict the existence of bright low-energy states and propose to extract the fine structure of the trion using the temperature dependence of emission spectra.

\vspace{6mm}

Corresponding author email: katarzyna.sadecka@pwr.edu.pl
\end{abstract}
\maketitle


\section{Introduction}

Gated bilayer graphene (BLG) is a voltage-tunable semiconductor whose properties have recently generated a great deal of experimental and theoretical interest \cite{zhang_direct_2009,wang_gate-variable_2008,ohta_controlling_2006,mak_observation_2009,castro_biased_2007}. While the low-energy electronic states of monolayer graphene are described by a massless Dirac fermion model \cite{geim_rise_2007,castro_electronic_2010,novoselov_two-dimensional_2005,trevisanutto_ab_2008}, the low-energy electron states of BLG are described by massive Dirac fermion model \cite{mccann_electronic_2013,zhang_direct_2009,wang_gate-variable_2008,ohta_controlling_2006,mak_observation_2009,castro_biased_2007,oostinga_gate-induced_2008}. Moreover, BLG is characterized by a pseudospin winding number of two and a valley-dependent Berry phase \cite{novoselov_unconventional_2006}, which leads to effects not observed in conventional semiconductors \cite{prada2021angular}. Furthermore, it has been demonstrated, both theoretically and experimentally, that BLG exhibits a continuously gate-tunable energy gap \cite{zhang_direct_2009,wang_gate-variable_2008,mccann_electronic_2013,min_ab_2007, aoki_dependence_2007,gava_ab_2009,falkovsky_gate-tunable_2010,ohta_controlling_2006,mak_observation_2009,castro_biased_2007,zhang-basov-prb2008,novoselov_unconventional_2006,ju_tunable_2017,sauer_exciton_2022}, allowing the study of new effects originating from the gap opening. This includes the gate dependence of interband transitions, including the presence of excitons \cite{wang_gate-variable_2008}, gate-tunable infrared phonon anomalies \cite{kuzmenko_gate_2009} or the gate-induced insulating states \cite{zhang_direct_2009,wang_gate-variable_2008,ohta_controlling_2006,mak_observation_2009,castro_biased_2007,oostinga_gate-induced_2008}. 

In parallel with gate controlled semiconductors, size controlled semiconductors and graphene QDs have been developed as building blocks of quantum technologies. Lasers, emitters, displays, detectors, and both single and entangled photon sources have been considered and built using self-assembled and graphene QDs confining both electrons and holes \cite{garcia2021semiconductor,bayer2000hidden,hawrylak2000excitonic,schwartz2016deterministic,laferriere2021systematic,Bester2003XinQD,hawrylak2014grapheneQD}. Simultaneously, laterally gated QDs allowed for the confinement of either electrons or holes, acting as single spin transistors and spin qubits, with potential application in quantum computing \cite{ciorga2000addition,hsieh2012physics,burkard2021semiconductor}. 
The confining potential in self-assembled and graphene QDs is determined by materials and structure, while the confining potential in gated QDs is determined by gate voltages and hence allows for very high tunability of their electronic properties. While laterally gated QDs are limited to confine either electrons or holes, self-assembled and graphene QDs allow to confine both carriers, however they are difficult to tune once they are made.

It has been shown that BLG QDs \cite{stampfer2023particle,Stampfer2020BLGQD,Stampfer2021BLGQD,Ensslin2019BLGQD,KorkusinskiBLGQD2023,SaleemExciton2023,pereira2007tunable,park_tunable_2010} can combine the ability to confine both electrons and holes of semiconductor self-assembled QDs with the tunability of laterally gated semiconductor QDs \cite{park_tunable_2010,SaleemExciton2023,KorkusinskiBLGQD2023}, thus enabling the existence of gate-tunable excitons in such nanostructures \cite{SaleemExciton2023}. In addition to excitons, the optical properties of semiconductor and graphene nanostructures are determined by trions \cite{Sheng2012GQD,Wojs1995TrionQD,Peeters1999TrionsQW,Peeters2001TrionsQW,Bayer2002TrionFineStructure,Witek2011TrionFineStructure,ozfidan_microscopic_2014,Ozfidan2015NanoLetters,Ozfidan2015Colloidal,Knothe2020BLGQD}.

In this work we present a theory of negatively charged excitons, i.e. trions \cite{Wojs1995TrionQD,Peeters1999TrionsQW,Peeters2001TrionsQW,Sheng2012GQD}, in a gated BLG QD. Following our previous work on tunable excitons \cite{SaleemExciton2023}, we begin with an atomistic approach to determine the energy levels of an electron and a hole laterally confined in the QD potential embedded in a BLG computational box containing up to $\sim$1.6 million atoms. We use an \textit{ab initio}-based tight-binding model described in section \ref{section:TightBinding}. In section \ref{section:TrionSpectrum} we derive the Bethe-Salpeter-like equation to describe a negatively charged exciton. We present a detailed study of the trion spectrum and show that, unlike in self-assembled QDs \cite{Bayer2002TrionFineStructure,Witek2011TrionFineStructure,Wojs1995TrionQD}, the trion in gated BLG QD contains a fine structure composed of ten low energy states related to valley and spin degrees of freedom. Similar degrees of freedom can be seen in colloidal graphene QDs \cite{ozfidan_microscopic_2014,hawrylak2014grapheneQD,Ozfidan2015Colloidal,Ozfidan2015NanoLetters} or in the \textit{p}-shell of self-assembled QDs. In section \ref{section:LightMatterInteraction} we present trion absorption and emission spectra. We observe that low-energy trion states are optically active. Furthermore, we propose a method of extracting the trion fine structure from finite temperature emission spectra.



\section{Tight-Binding Model}
\label{section:TightBinding}

\subsection{Gated Bilayer Graphene}

\begin{figure}[t]
    \centering
    \includegraphics[width=\linewidth]{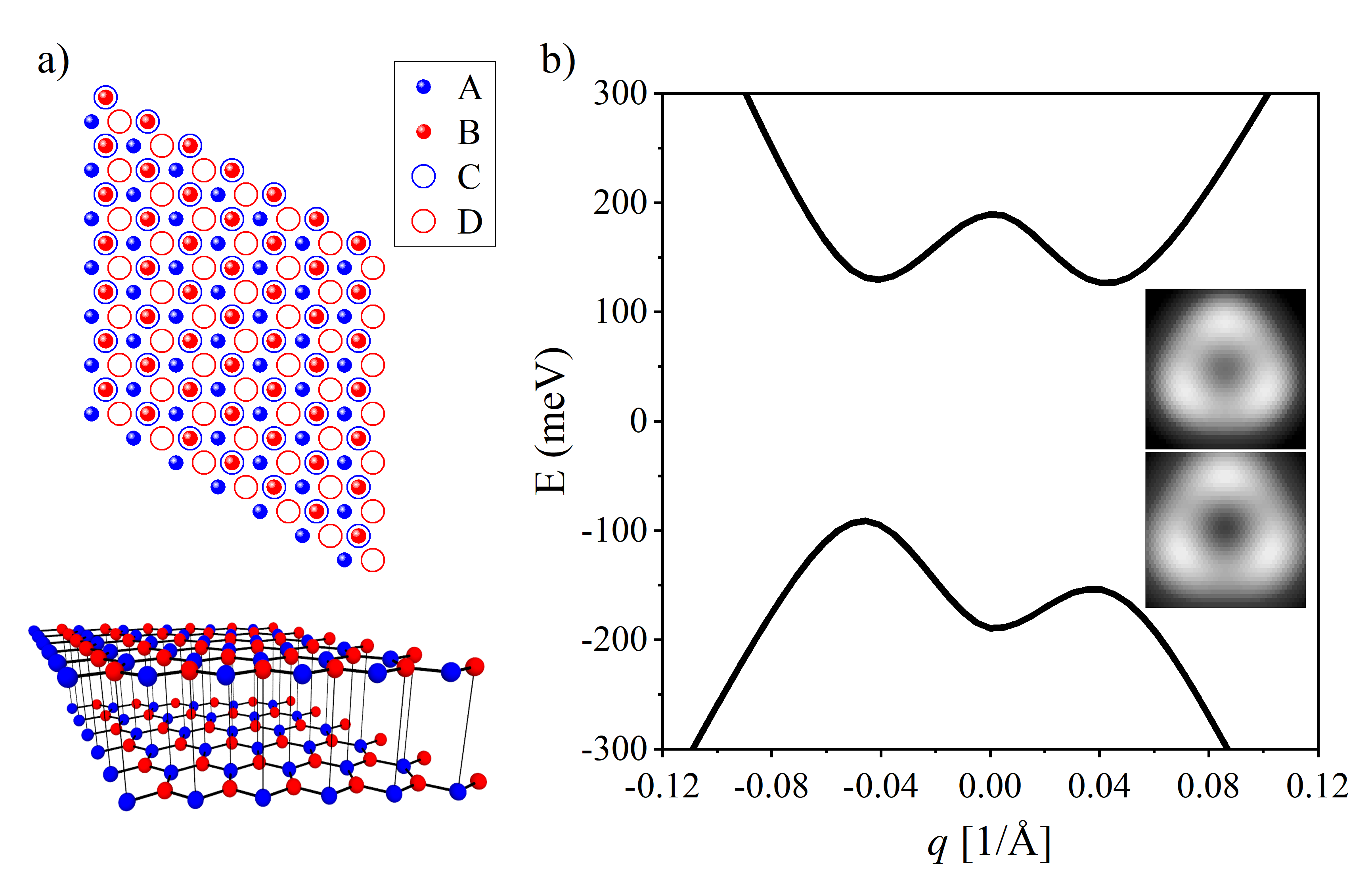}
    \caption{{ Electronic properties of gated Bernal-stacked bilayer graphene.} (a) The upper plot shows the BLG rhomboidal computational box in a schematic way allowing to distinguish sublattices A, B (layer 1, shown with full symbols) and C, D (layer 2, shown with circles). The bottom scheme presents a 3-dimentional (3D) view of the BLG in the Bernal-stack layer configuration. (b) The bulk electronic structure of BLG in the vicinity of the K point with the displacement voltage $V_E=0.38$ eV. $q$ describes the distance from K point along the $\vec{b_1}+\vec{b_2}$ direction, where $\vec{b_1}$ and $\vec{b_2}$ are the reciprocal space lattice vectors. The insets present the energy heatmaps of conduction (upper plot) and valence (bottom plot) band in the close vicinity of K point.}
   \label{fig:system}
\end{figure}

Following our previous works \cite{SaleemExciton2023,KorkusinskiBLGQD2023} we consider the Bernal-stacked BLG, as presented in Fig.~\ref{fig:system}(a), where the 4 sublattices are denoted as A, B (layer $1$) and C, D (layer $2$). The in-plane bond length is $a=0.143$ nm, and the distance between layers is $h=0.335$ nm. We have chosen the unit cell vectors as $\vec{a}_1=a\left(0,\sqrt{3}\right)$ and $\vec{a}_2=\frac{a}{2}\left(3,-\sqrt{3}\right)$. The rhomboidal real-space computational box shown in a schematic way in Fig.~\ref{fig:system}(a) has been generated by the vector $\vec{R}_{m_1,m_2}=m_1\vec{a}_1+m_2\vec{a}_2$, where $m_1=\{-\left(N_1-1\right)/2,...,\left(N_1-1\right)/2$\} and $m_2=\{-\left(N_2-1\right)/2,...,\left(N_2-1\right)/2\}$, with $N_1=N_2$ being the number of unit cells along the directions of the lattice vectors. With $N_1=N_2=633$ used here, this creates a computational box containing about 1.6 million carbon atoms, making the problem computationally demanding. In order to remove the finite-size effects we impose periodic boundary conditions connecting the opposite edges of the rhombus and giving us a set of allowed $\vec{k}$-vectors  \cite{SaleemExciton2023,KorkusinskiBLGQD2023}. There are 4 carbon atoms in each unit cell generating 4 sublattices. For each sublattice $l$ we build a Bloch function as $\ket{\phi_{\vec{k}}^l}=\frac{1}{\sqrt{N_1N_2}}\sum_{\vec{R}_l}e^{i\Vec{k}\cdot \vec{R}_l}\ket{\vec{R}_l}$, where $\ket{\vec{R}_l}$ is a $p_z$ Slater orbital localized at a position $\Vec{R}_l$. In the next step we apply an external electric field perpendicular to the graphene layers, with the applied potential $+V_E/2$ on layer 1 and $-V_E/2$ on layer 2, giving the potential difference between layers $V_E$. Our Hamiltonian in the basis of sublattices A, B, C, D is given by:

\begin{equation}
    H\left(\vec{k}\right)  = \left(
    \begin{array}{cccc}
         \frac{V_E}{2} & \gamma_0 f\left(\vec{k}\right) & \gamma_4 f\left(\vec{k}\right) & \gamma_3 f^*\left(\vec{k}\right)  \\
         \gamma_0 f^*\left(\vec{k}\right) & \frac{V_E}{2} & \gamma_1 & \gamma_4 f\left(\vec{k}\right)    \\
         \gamma_4 f^*\left(\vec{k}\right) & \gamma_1 & -\frac{V_E}{2} & \gamma_0 f\left(\vec{k}\right) \\
         \gamma_3 f\left(\vec{k}\right) & \gamma_4 f^*\left(\vec{k}\right) & \gamma_0 f^*\left(\vec{k}\right) & -\frac{V_E}{2}
    \end{array}
    \right),
    \label{eq:BulkH}
\end{equation}
where $\gamma_0=-2.5$ eV defines the nearest neighbour (NN) intralayer hopping and $\gamma_1=0.34$ eV defines the interlayer hopping between the AB stacked atoms. The displacement voltage in this example is $V_E=0.38$ eV. We include the effects of trigonal warping via the matrix element $\gamma_3=0.12\left|\gamma_0\right|$ eV \cite{KorkusinskiBLGQD2023,Knothe2020BLGQD}, and we take $\gamma_4$ as equal to $\gamma_3$ \cite{KorkusinskiBLGQD2023}. The function $f\left(\vec{k}\right)$ is defined as 
$f\left(\vec{k}\right)=e^{i\vec{k}\cdot\vec{\tau}}  \left(  1 + e^{-i\vec{k}\cdot \left( \vec{a_1}+\vec{a_2}\right)} + e^{-i\vec{k}\cdot\vec{a_2}}\right)$,
where $\vec{\tau}=\left(a,0\right)$. The electron wavefunction for a given band $p$ is a linear combination of Bloch wavefunctions of each sublattice, $\Phi_{p,\vec{k}}\left(\vec{r}\right) = \sum_l\mu_{p,\vec{k}}^l\phi_{\vec{k}}^l\left(\vec{r}\right)$, where $l$ denotes the 4 sublattices, A, B, C and D, respectively. The coefficients $\mu_{p,\vec{k}}^l$ and energies $\varepsilon_{p,\vec{k}}$ are obtained by diagonalizing the Hamiltonian shown in Eq.~(\ref{eq:BulkH}).

Fig.~\ref{fig:system}(b) shows the energy spectrum $\varepsilon_{p,\vec{k}}$ for the two lowest energy bands, the topmost valence band and the bottom conduction band. While in the absence of electric field $(V_E=0)$ the valence band (VB) and the conduction band (CB) touch at the K point, applying the vertical electric field results in opening the energy gap of BLG. The electron-hole symmetry is broken due to the inclusion of non-zero matrix elements $\gamma_4$, which is manifested in the difference between the effective mass of an electron and a hole around the K valley, as depicted in Fig.~\ref{fig:system}(b). Moreover, the introduction of trigonal warping has reduced the symmetry of the system from C$_\infty$ to C$_3$ \cite{henriques2022absorption}, giving rise to the three energetic maxima/minima in the VB / CB, which are shown in the insets of Fig.~\ref{fig:system}(b). However, the electronic structure still resembles the characteristic Mexican hat-like dispersion around the K and K’ points of the Brillouin zone. Near the K and K’ points, the wavefunction in the CB and VB differ by a phase of $2\phi_k$ on the two highly occupied sublattices A, D. This is a consequence of the winding number of 2 in BLG \cite{pereira2007tunable}.

\subsection{Gate-Defined Quantum Dot}

We showed in Ref.~\onlinecite{SaleemExciton2023} that by applying a lateral potential, modeled by an \textit{ab initio}-based screened Gaussian potential $V^{eff}_{QD}$, we are able to confine both electrons and holes in a bilayer graphene quantum dot. This potential is given as 
\begin{equation}
    \hat{V}^{eff}_{QD}\left(\rho\right)=
    \begin{cases} 
        - c_1 e^{\left(-\frac{\alpha_1\rho^2}{r^2_{QD}}\right)} - c_2 e^{\left(-\frac{\alpha_2\rho^2}{r^2_{QD}}\right)}, & z= 0,\\
        + c_1 e^{\left(-\frac{\alpha_1\rho^2}{r^2_{QD}}\right)} + c_2 e^{\left(-\frac{\alpha_2\rho^2}{r^2_{QD}}\right)}, & z=-h ,
   \end{cases}
   \label{eq:ScreenedPotential}
\end{equation}
with parameters $c_1=-0.018$ eV, $c_2=0.207$ eV, $\alpha_1=6.128$, and $\alpha_2=1.006$ for the QD radius $r_{QD}=20$ nm.

The QD electronic states are expanded in the basis of band states of the computational box as 
$\varphi_s(\vec{r}) = \sum_p\sum_{\vec{k}} \nu^s_{p,\vec{k}} \Phi_{p,\vec{k}}\left(\vec{r}\right)$,
where the summation is carried over the 4 bands denoted by $p$ and wavevectors $\vec{k}$ defined by the computational rhombus. Solving the Schr\"odinger equation results in the eigenequation for the amplitudes $\nu^s_{p,\vec{k}}$:
\begin{equation}
    \varepsilon_{p,\vec{k}} \nu^s_{p,\vec{k}} + \sum_{p',\vec{k'}} \braket{\Phi_{p,\vec{k}}|\hat{V}^{eff}_{QD}|\Phi_{p',\vec{k'}}} \nu^s_{p',\vec{k'}} = \epsilon_s \nu^s_{p,\vec{k}},
    \label{eq:QDequation}
\end{equation}
where $\varepsilon_{p,\vec{k}}$ denotes the eigenvalues of the bulk Hamiltonian described by Eq.~(\ref{eq:BulkH}). 

We note that the QD potential confining electrons in the QDs consists of two contributions, an electrostatic contribution and a band contribution. The matrix elements coupling the band state is given by
\begin{equation}
     \braket{\Phi^{p}_{\vec{k}}|V^{eff}_{QD}|\Phi^{p'}_{\vec{k}'}} =\sum_l \mu_{p',\vec{k'}}^{l} \left(\mu_{p,\vec{k}}^{l}\right)^* e^{i\left(\vec{k}'-\vec{k}\right)\cdot d_l} V^{eff}_{\vec{k},\vec{k}',l},
    \label{eq:QDpotential}
\end{equation}
where $d_{l}$ denotes the position of atoms within a unit cell, and $V^{eff}_{\vec{k},\vec{k}',l}$ is the Fourier transform of the QD potential $\hat{V}^{eff}_{QD}$ on sublattice $l$.

Solving Eq.~(\ref{eq:QDequation}) generates the QD energy spectrum. Following the methodology from our previous works \cite{SaleemExciton2023,KorkusinskiBLGQD2023} we limit our calculations to the band states within a certain energy cutoff 
$-E_{cut}\leq\varepsilon_{p,\vec{k}}\leq E_{cut}$. We take $E_{cut}=600$~meV, which is larger than the energy scale presented in Fig.~\ref{fig:system}(b). This energy cutoff defines our basis as states in a close vicinity of K and K$'$ points. 

\begin{figure}[t]
    \includegraphics[width=\linewidth]{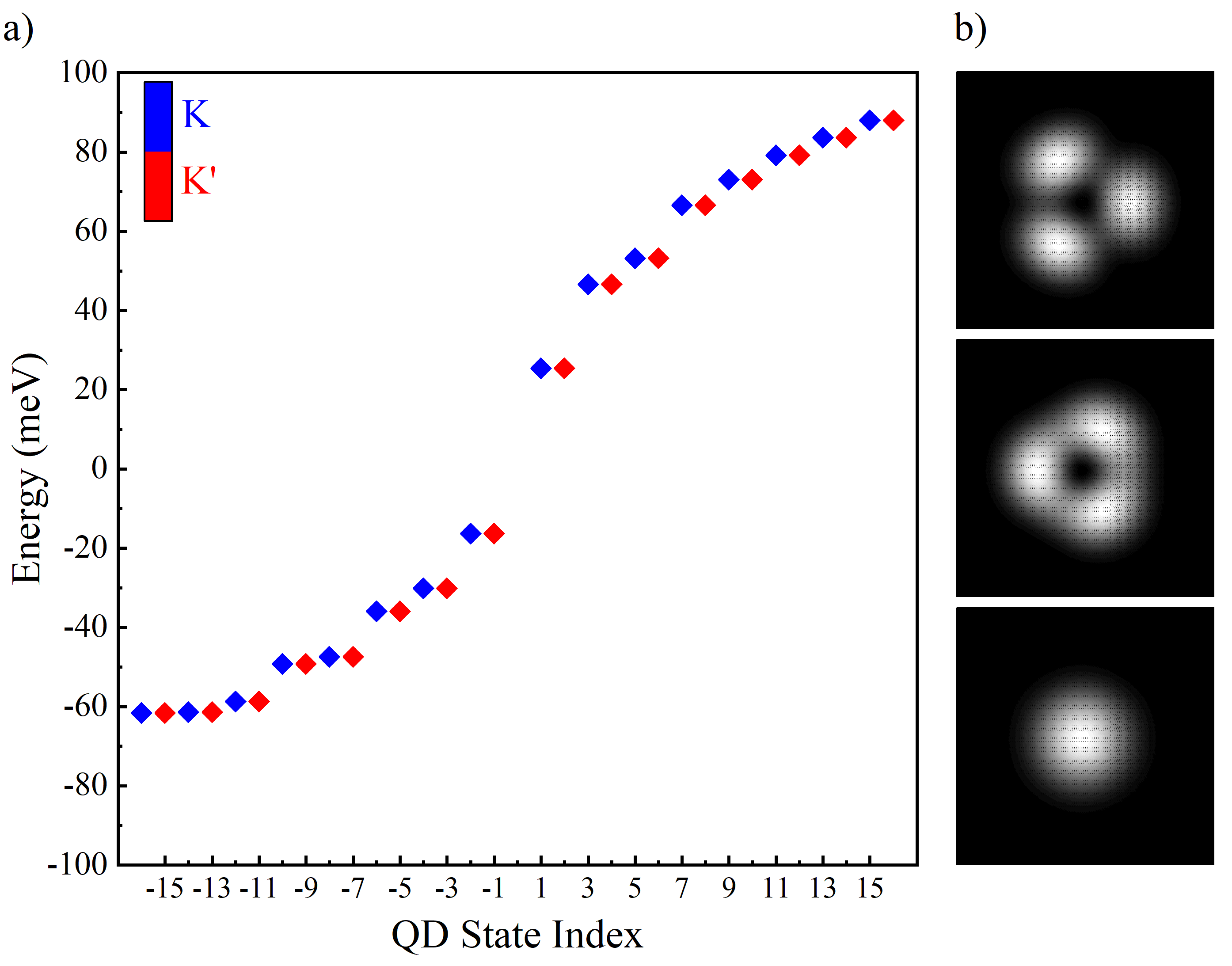}
    \caption{{ Bilayer graphene quantum dot spectrum.} (a) Valley-resolved single-particle energy spectrum of gated BLG QD with a radius $r_{QD}=20$ nm. 32 QD states around the energy gap have been presented, where negative indices describe valence band states, while positive indices correspond to conduction band states. The valley degree of freedom has been detoned with color (blue describes K valley, and red denotes K$'$ valley). (b) Charge density distribution of the first 3 CB shells for the dominantly occupied (top) layer, ordered starting from the bottom to the top, where the first graph is corresponding to the most bottom CB state (indexed by 1).}
   \label{fig:SPQDEnergies}
\end{figure}

Figure~\ref{fig:SPQDEnergies}(a) shows the QD energy spectrum for a QD with the radius $r_{QD}=20$ nm. We present the spectra for both K and K$'$ valley (denoted by blue and red, respectively) around the Fermi level, where odd/even numbers are indexing states from K/K$'$ valley. Each state is additionally doubly-degenerate due to spin. Similar to the bulk dispersion, the electron-hole symmetry in the QD is broken as a result of introducing $\gamma_4$ in the bulk Hamiltonian~(\ref{eq:BulkH}). 
Moreover, the QD energy spectrum appears to be different from the characteristic quantum dot spectra for gated lateral QD in GaAs or self-assembled QDs showing electronic shells of a 2D harmonic oscillator) \cite{hawrylak1993single,gomes2021variational,raymond2004excitonic,bayer2000hidden}. The charge density distribution on the top layer of the lowest energy CB shells for valley K is presented in Fig.~\ref{fig:SPQDEnergies}(b). We see that the lowest energy CB doublet is \textit{s}-like, followed by two \textit{p}-like shells, which are split in energy. The splitting originates from the topological properties of BLG and the emergence of the winding number of 2 near the K point \cite{pereira2007tunable}. However, since the inclusion of trigonal warping has reduced the symmetry of the system, we now observe the reduction of charge density symmetry from C$_\infty$ to C$_3$.



\section{Trion Spectrum}
\label{section:TrionSpectrum}

\begin{figure*}[t]
    \centering
     \includegraphics[width=1.0\textwidth]{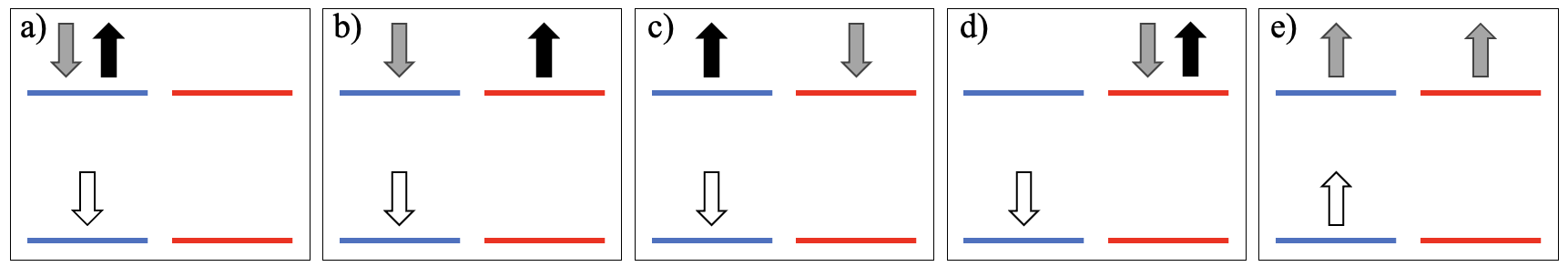}
    \caption{{} Diagram representation of the low-energy trion configurations. Vertical arrows represent carriers (electrons/holes are denoted by full/empty arrows) characterized by spin (arrow pointing up/down denotes spin up/down). The states are built only from the hole residing in the bottom of the VB and two electrons in the bottom of the CB. The grey-coloured arrow denotes the electron with spin corresponding to the spin of the hole. The colour of the horizontal lines denotes valley (blue/red corresponds to K/K$'$ valley).}
   \label{fig:Configurations}
\end{figure*}

In this section we include the analysis of electron-electron interactions. The many-electron Hamiltonian written in the basis of QD single-particle states has the following form:
\begin{equation}
\begin{split}
    \hat{H}_{MB} &= \sum_{p,\sigma} \epsilon_p c_{p,\sigma}^\dagger c_{p,\sigma} \\
              &+ \frac{1}{2} \sum_{p,q,r,s}\sum_{\sigma,\sigma'} \braket{p,q|V_C|r,s} c_{p,\sigma}^\dagger c_{q,\sigma'}^\dagger c_{r,\sigma'} c_{s,\sigma} \\
              &- \sum_{p,s,\sigma} V_{p,s}^{P} c_{p,\sigma}^\dagger c_{s,\sigma},
\end{split}
\label{eq:mbHamiltonian} 
\end{equation}
where $c^\dagger_{p,\sigma}/c_{p,\sigma}$ creates/annihilates an electron with spin $\sigma$ in a QD state $p$. Here, the indices $p,q,r,s$ run over all QD states. The electron-electron interactions $V_C$ are taken as Coulomb interactions screened by the dielectric constant $\kappa$, $V_C=\frac{e^2}{\kappa|\vec{r}_1-\vec{r}_2|}$, with $\kappa=3.9$ \cite{Zarenia2013kappa,Dean2010BNsubstrate}. The Coulomb matrix elements $\braket{p,q|V_C|r,s}$ can be expressed in terms of atomic orbitals and computed numerically with Slater-like $p_z$ orbitals (for computational details see Ref.~\onlinecite{KorkusinskiBLGQD2023}). The last term of Eq.~(\ref{eq:mbHamiltonian}) accounts for the positive charge of the background characterized by the same charge distribution as that of electrons in the fully occupied VB. It has been introduced in order to ensure the overall charge neutrality and is defined as $V_{p,s}^{P}=2\sum_m^{N_{occ.}}\braket{p,m|V_C|m,s}$.

For the study of many-body effects we have approximated the many-electron ground state at half-filling as a single Slater determinant of all the occupied VB states $\ket{GS}=\prod_{p,\sigma}c^\dagger_{p,\sigma}\ket{0}$. 
We now construct the negatively charged trion states $\psi_-^n$ as a linear combination of 3-body states containing one electron-hole pair excitation and one additional electron in the CB, 
$\ket{\alpha,\sigma';\beta,\sigma;i,\sigma}=c^\dagger_{\alpha,\sigma'}c^\dagger_{\beta,\sigma}c_{i,\sigma} \ket{GS}$.
Here the hole is defined as the lack of electron in the VB, $c_{i,\sigma}\ket{GS}$, respectively.
We use Greek letters to denote electrons in CB, and Latin letters for holes, missing electrons, in 
the VB. Because of the two non-equivalent valleys in BLG QDs we can construct ten low-energy trion configurations with $S_z=\frac{1}{2}$ , constructed from states lying at the top of the VB and at the bottom of the CB. Fig.~\ref{fig:Configurations} shows five of these configurations; the other five configurations correspond to moving the hole from valley K (blue) to valley K$'$ (red). In the simplest configuration, one can expect an intravalley trion shown in Fig.~\ref{fig:Configurations}(a), where both CB electrons and the hole reside in the same valley (either K or K$'$), and the spins of the electrons are opposite. Another class of configurations involves an exciton in one valley and additional electron in the opposite valley creating configurations depicted in Fig.~\ref{fig:Configurations}(b) and (c), respectively. In this case one observes an intravalley exciton correlated with an excess electron residing in the opposite valley. Furthermore, both electrons can be placed in the same valley, while the hole resides in the opposite valley, as presented in \ref{fig:Configurations}(d). Finally, one can observe a trion configuration where all carriers are characterized by the same spin. This situation has been depicted in Fig.~\ref{fig:Configurations}(e).

The trion wavefunction is a linear combination of all possible trion configurations and reads as follows:
\begin{equation}
    \ket{\psi_-^n} = 
        \sum_{\alpha,\beta,i}\sum_{\sigma,\sigma'} A^n_{\alpha,\beta,i,\sigma,\sigma'}
        c^\dagger_{\alpha,\sigma'}c^\dagger_{\beta,\sigma}c_{i,\sigma} \ket{GS}.
    \label{eq:trionWF}
\end{equation}
In the above definition, $A^n_{\alpha,\beta,i,\sigma,\sigma'}$ is the amplitude of the 3-body configuration for a trion state $n$. The Greek indices $\alpha$, $\beta$ in the sums run over all QD conduction band states, while the Latin index $i$ goes through all QD valence band states. For $\sigma=\sigma'$ only the configurations with $\alpha>\beta$ have to be considered in order to avoid double counting of configurations. In the further analysis we will hide the spin index in the Greek and Latin letters.

\begin{figure*}[t]
     \includegraphics[width=1.0\textwidth]{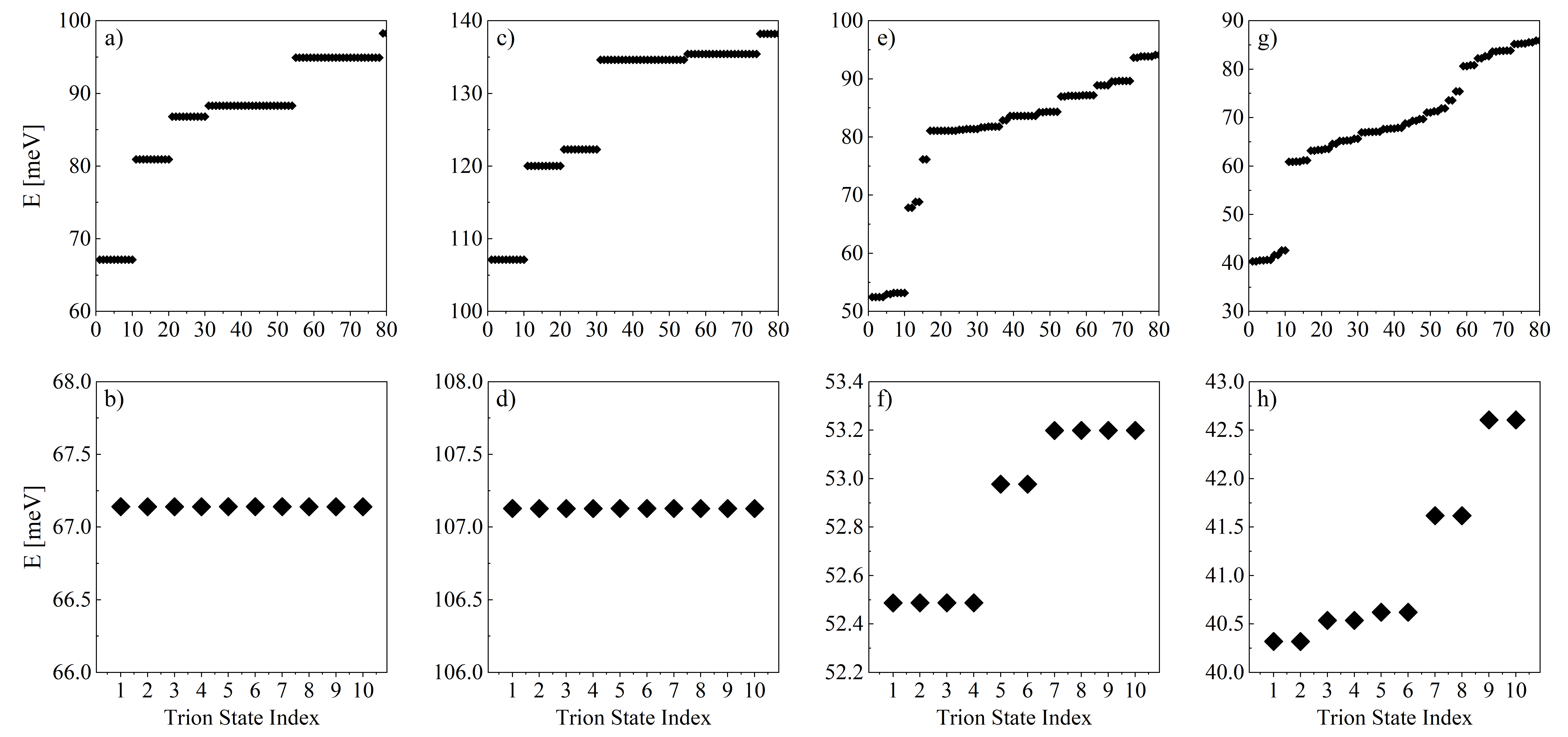}
    \caption{{Trion energy spectra presented in different stages of including Coulomb interactions.} (a,b) Energy levels for non-interacting trion states. (c,d) Trion spectrum renormalized by self-energies. (e,f) Energy spectrum accounting for the self-energies and the electron-hole direct and exchange interaction. (g,h) Spectrum obtained as a full BSE solution. Top row, (a,c,e,g), presents energy levels for the first 80 trion states obtained as a solution of the Bethe-Salpeter Eq.~(\ref{eq:BSE}). Bottom row, (b,d,f,h), shows zooms on the low energy manifold (first 10 trion energy states) presenting the trion fine structure.}
   \label{fig:TrionSpectrum}
\end{figure*}

The many body Hamiltonian renormalizes the energy of quasi-electrons and quasi-holes and mixes different configurations. As a result, the amplitudes $A^n_{\alpha,\beta,i}$ of the trion states satisfy the Bethe-Salpeter-like equation (BSE) \cite{DeilmannTrion2016,TorcheBesterTrion2019,DeilmannTrionErratum,Wojs1995TrionQD}:
\begin{equation}
\begin{split}
    E^n A^n_{\alpha,\beta,i} &= 
    \left( \epsilon_\alpha + \epsilon_\beta - \epsilon_i \right) A^n_{\alpha,\beta,i} \\
    &+\sum_{\mu,\lambda,j} \left( 
    \begin{array}{c} 
    +\Sigma_{\mu,\alpha}\delta_{i,j}\delta_{\beta,\lambda}  
    -\Sigma_{\lambda,\alpha}\delta_{i,j}\delta_{\beta,\mu} \\
    +\Sigma_{\lambda,\beta}\delta_{i,j}\delta_{\alpha,\mu}  
    -\Sigma_{\mu,\beta}\delta_{i,j}\delta_{\alpha,\lambda} \\
    -\Sigma_{i,j}\delta_{\alpha,\mu}\delta_{\beta,\lambda}
    \end{array}
    \right) A^n_{\mu,\lambda,j} \\
    &+\sum_{\mu,\lambda,j} \left[
    \begin{array}{c}
        +\left( V_{i,\lambda,j,\beta}-V_{i,\lambda,\beta,j} \right)\delta_{\alpha,\mu} \\
        +\left( V_{i,\mu,j,\alpha}-V_{i,\mu,\alpha,j}\right)\delta_{\beta,\lambda} \\
        -\left( V_{i,\lambda,j,\alpha}-V_{i,\lambda,\alpha,j}\right)\delta_{\beta,\mu} \\
        -\left( V_{i,\mu,j,\beta}-V_{i,\mu,\beta,j} \right)\delta_{\alpha,\lambda} \\
        -\left( V_{\mu,\lambda,\alpha,\beta} - V_{\mu,\lambda,\beta,\alpha}\right)\delta_{i,j}
    \end{array}
    \right] A^n_{\mu,\lambda,j},
\end{split}
\label{eq:BSE}
\end{equation}
where $\epsilon_p$ denotes the QD single-particle energies, $\Sigma_{p,q}=-\sum_m^{N_{occ.}}\braket{m,p|V_C|m,q}$ defines the scatterings containing the self-energy terms $\Sigma_{p,p}$, $V_{p,q,r,s}=\bra{p,q}V_C\ket{r,s}$ corresponds to the Coulomb matrix elements,and $N_{occ.}$ corresponds to the number of occupied states.

Let us now focus on the energy of trion configurations, the diagonal part of the Bethe-Salpeter matrix: 
\begin{equation}
\begin{split}
    \bra{i,\beta,\alpha}&\hat{H}_{MB}\ket{\alpha,\beta,i} = \\
    &\left( \epsilon_\alpha + \Sigma_{\alpha,\alpha} \right) +
    \left( \epsilon_\beta + \Sigma_{\beta,\beta} \right) -
    \left( \epsilon_i + \Sigma_{i,i} \right) \\
    &- V_{i,\alpha,\alpha,i} + V_{i,\alpha,i,\alpha}
    - V_{i,\beta,\beta,i} + V_{i,\beta,i,\beta} \\
    &+ V_{\alpha,\beta,\beta,\alpha} - V_{\alpha,\beta,\alpha,\beta}.
\end{split}
\label{eq:BSEdiagonal}
\end{equation}
It contains 12 terms. 
The terms 
$\left( \epsilon_\alpha + \Sigma_{\alpha,\alpha} \right) + \left( \epsilon_\beta + \Sigma_{\beta,\beta} \right) - \left( \epsilon_i + \Sigma_{i,i} \right)$ show the contribution from 3 QD single-particle energies $\epsilon_p$ (one for hole in the VB and two for electrons in the CB) corrected by its exchange self-energies $\Sigma_{p,p}$. 
The next part of Eq.~(\ref{eq:BSEdiagonal}), 
$-V_{i,\alpha,\alpha,i} + V_{i,\alpha,i,\alpha} - V_{i,\beta,\beta,i} + V_{i,\beta,i,\beta}$, 
contains 6 Coulomb scattering terms, 3 of them are direct ($V_{p,q,q,p}$) and 3 are exchange ($V_{p,q,p,q}$). The terms $V_{i,\alpha,\alpha,i}$ and $V_{i,\alpha,i,\alpha}$ describe respectively the direct and exchange interaction of electron $\alpha$ with the hole $i$, while $V_{i,\beta,\beta,i}$ and $V_{i,\beta,i,\beta}$ correspond respectively to the direct and exchange interaction of electron $\beta$ with the hole $i$. The last two Coulomb matrix elements, $V_{\alpha,\beta,\beta,\alpha}$ and $V_{\alpha,\beta,\alpha,\beta}$, correspond respectively to the direct and exchange interaction between the electrons $\alpha$ and $\beta$ in the CB.

We obtain the trion spectrum by solving Eq.~(\ref{eq:BSE}) in the subspace of 32 single-particle states around the Fermi level, as presented in Fig.~\ref{fig:SPQDEnergies}. We restrict the trion states to the $S_z=+\frac{1}{2}$ subspace. The self-energies $\Sigma_{p,p}$ have converged for $N_{occ.}=120$ filled valence band QD states \cite{SaleemExciton2023}. 
 
We now move to a detailed description of the calculated trion spectrum, where we study the evolution of the spectrum through the inclusion of different types of Coulomb interactions. We note that each state is at least doubly degenerate due to valley. Thus we restrict our analysis to states in which the hole resides in the valley K, as depicted in Fig.~\ref{fig:Configurations}. Fig.~\ref{fig:TrionSpectrum} shows the spectra obtained by solving Eq.~(\ref{eq:BSE}). With 32 single particle levels this involves 6016 trion configurations. Each column corresponds to a different stage of including Coulomb interactions. Furthermore, the bottom row shows zooms into the low-energy manifold of states.

We start the analysis by computing only the non-interacting trion spectrum, when $V_C\equiv 0$. The result is presented in Fig.~\ref{fig:TrionSpectrum}(a,b). We find a degenerate low-energy manifold of 10 trion states, corresponding to the 5 negatively charged exciton configurations presented schematically in Fig.~\ref{fig:Configurations} and their valley-symmetric equivalents. These states are built only from the hole residing in the top VB and two electrons in the bottom CB. The trion spectrum for higher energies is more complex due to the possible trion configurations and broken electron-hole symmetry.

In the next step we include the self-energy correction of the electron and hole energies, $\Sigma_{p,q}$. The resulting trion spectrum is shown in Fig.~\ref{fig:TrionSpectrum}(c) and (d). We observe a large blueshift of the trion energies. Furthermore, the self-energies introduce a reordering of the high-energy manifolds. This effect originates from the difference between the electron and hole self-energy.
 
In Fig.~\ref{fig:TrionSpectrum}(e,f) we present the trion spectrum obtained by including in Eq.~\ref{eq:BSE} the self-energy terms $\Sigma_{p,q}$ and diagonal parts of the Coulomb direct and exchange interactions, $V_{p,q,q,p}=\bra{p,q}V_C\ket{q,p}$ and $V_{p,q,p,q}=\bra{p,q}V_C\ket{p,q}$, respectively. While the direct electron-electron and electron-hole interaction do not depend on valleys, the exchange interaction is different within the same valley and between different valleys. Moreover, the electron-electron interaction magnitude is different from that of the electron-hole interaction. The electron-hole intravalley exchange interaction is large, $\sim O\left(10^{-1}\right)$~meV, and is responsible for raising the energy of the electron-electron-hole complex. The relatively smaller electron-electron exchange interaction, $\sim O\left(10^{-2}\right)$~meV, is repulsive and lowers the trion energy. The last type of exchange interaction, that also raises the energy of the negatively charged exciton, is the electron-hole intervalley exchange interaction. However, this interaction is the smallest, $\sim O\left(10^{-3}\right)$~meV in this example, and can be neglected in the analysis of the low-energy manifold. These exchange elements result in the splittings of the low-energy manifold. The first group is four-fold degenerate. The two ''valley unique'' states correspond to the configurations (c) and (d) in Fig.~\ref{fig:Configurations}. These configurations do not allow for intravalley electron-hole exchange, unlike the other 3 configurations corresponding to (a), (b) and (e) in Fig.~\ref{fig:Configurations}, and as such are the lowest energy states. Next we find a doublet corresponding to the spin $\left(\uparrow\uparrow\uparrow\right)$ configurations, as shown in Fig.~\ref{fig:Configurations}(e). These configurations allow for electron-electron exchange and thus lower the energy relative to the four states above it. The last low-energy shell is again four-fold degenerate, where the electron and hole with the same spin belong to the same valley. This corresponds to the configurations (a) and (b) in Fig.~\ref{fig:Configurations}. These configurations contain intravalley electron-hole exchange, but their energy is not lowered by electron-electron exchange, making them the highest energy states in the manifold of 10 states. We also observe reordering of higher-energy states due to large exchange splittings compared to the single-particle level spacing, as presented in Fig.~\ref{fig:TrionSpectrum}(e).

Finally, we include the effects of correlations. The resulting trion spectrum is presented in Fig.~\ref{fig:TrionSpectrum}(g,h). In Fig.~\ref{fig:TrionSpectrum}(g) the first 80 correlated trion states have been shown. We observe a manifold of 10 low-energy trion states, separated from the higher energy states by a large gap. Each trion state comes as a doublet due to valley degeneracy. These states are mainly built from the 10 possible low-energy 3-body configurations, where the hole occupies the top of the VB, and the electrons are located in the bottom of the CB. A zoom into the first 10 trion states has been presented in Fig.~\ref{fig:TrionSpectrum}(h). Unlike in gated lateral QDs in GaAs or self-assembled QDs, where trions do not contain any fine structure beyond spin \cite{Bayer2002TrionFineStructure,Witek2011TrionFineStructure,Ediger2007}, we observe a rich low-energy manifold of trion states \cite{narvaez2005excitons,torche2019first}. In the absence of Coulomb interaction, the low-energy states of the trion are exactly degenerate. With the inclusion of electron-hole and electron-electron exchange interaction, the states split into 3 groups. When including the effects of correlations, the low-energy manifold forms 5 groups of doublets. The splittings are proportional to intravalley and intervalley exchange between the 2 electron-hole complexes and the electron-electron complex. The splittings are further corrected by correlations. The lowest energy doublet is a spin $\frac{3}{2}$ state, one formed in the K valley and another in K$'$. The 8 states at higher energy are spin $\frac{1}{2}$ states. Finally we determine the trion binding energy to be $E^T_b\approx-6.2$~meV (see Appendix).



\begin{figure}[t]
    \centering
    \includegraphics[width=\linewidth]{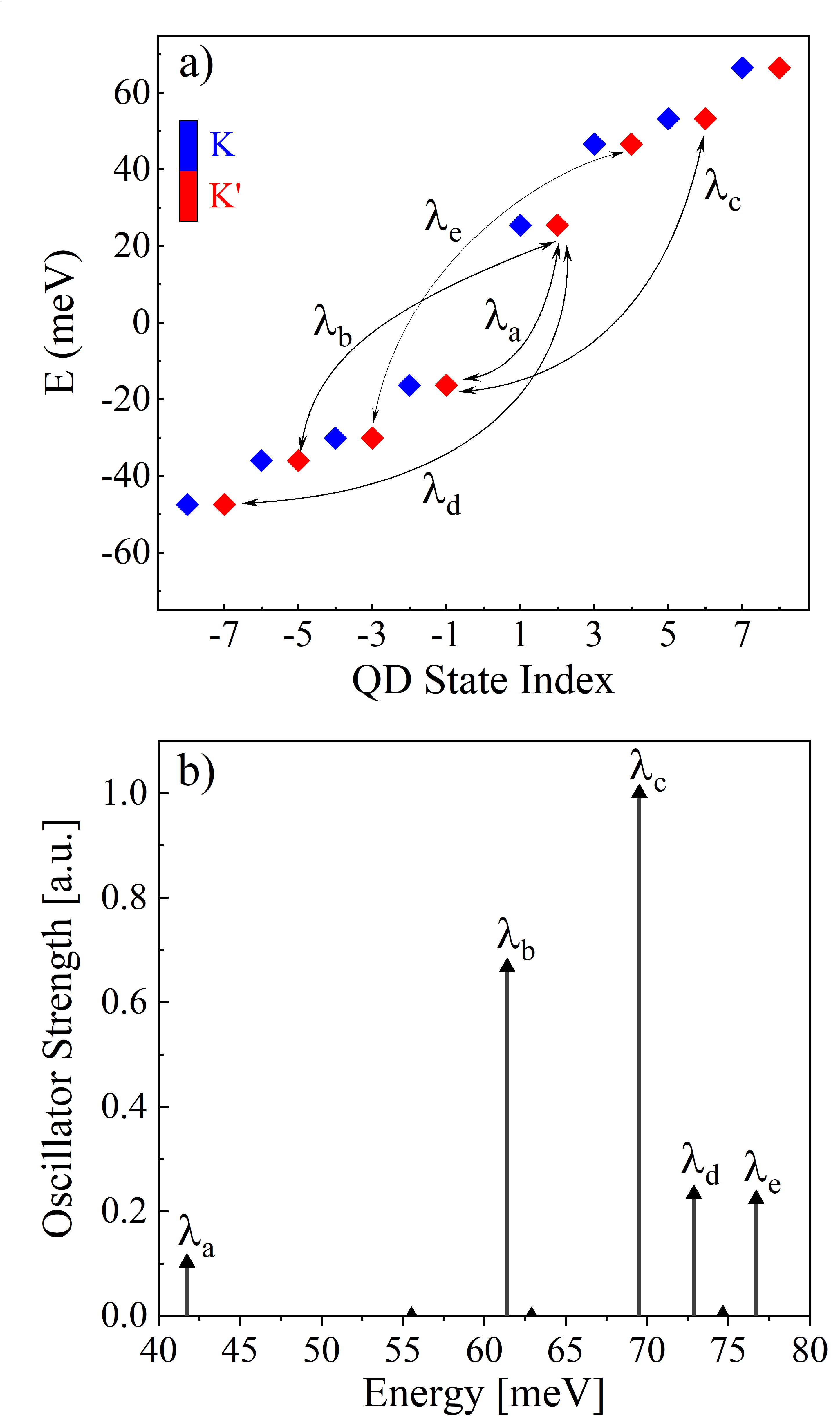}
    \caption{{Dipole moments for optical transitions.} 
    (a) Valley-resolved single-particle energy spectrum of gated BLG QD for 16 QD states around the energy gap. Blue denotes K valley, and red denotes K$'$ valley, respectively. Black arrows connect the VB and CB characterised by the largest dipole matrix elements. (b) Joint optical density of states for the non-interacting electron-hole pair. Black arrows show the magnitude of the corresponding dipole elements.}
   \label{fig:DipoleMoments}
\end{figure}

\section{Light-Matter Interaction}
\label{section:LightMatterInteraction}

In this section we describe the interaction of the trion in BLG QD with light. We look at the influence of the excess carrier on the absorption and emission spectra given by Fermi's golden rule. Bilayer graphene quantum dot couples with light via dipole matrix elements (DME) 
$D_{p,q}=\bra{\varphi_p}\mathcal{E}^{\pm}_0\cdot\vec{r}\ket{\varphi_q}$ 
between QD states $p$ and $q$, where $\vec{r}$ is the position operator and $\mathcal{E}^{\pm}_0$ is the electric field of the circularly polarized light. In Fig.~\ref{fig:DipoleMoments}(a) we show which QD states are coupled by DME. We connect with arrows the VB states connected with the CB states by large dipole elements $\vec{D}_{p,q}$. We notice that, unlike in self-assembled QDs, the top VB state (labeled by the QD index $-1$) is not strongly connected with the lowest energy CB state, but rather with the third shell ($+5$) in the same valley. We denote this transition as $\lambda_a$. A symmetric transition, $\lambda_b$, connects the state at the bottom of CB ($+1$) with the third VB shell ($-5$). It is worth noting that all intervalley transitions are characterised by negligibly small DME. Moreover, similar to TMDs monolayers, only one valley is optically active for a given circular polarization of light \cite{Szulakowska2019MagnetoX,Xu2014TMDreview}.

In Fig.~\ref{fig:DipoleMoments}(b) we present the joint optical density of states for non-interacting electron-hole pairs. We label the peaks according to the transitions presented in the Fig.~\ref{fig:DipoleMoments}(a). Similar to our previous work \cite{SaleemExciton2023}, we observe that the brightest peak $\lambda_c$ corresponds to the transition from the top VB shell ($-1$) to the third CB shell ($+5$). The transition from the top VB shell ($-1$) to the bottom CB shell ($+1$) is not forbidden , but characterized by low oscillator strength. This originates from the inclusion of trigonal warping in the bulk Hamiltonian described by Eq.~(\ref{eq:BulkH}). This complex light-matter coupling is furthermore significantly modified by the electron-electron interactions.

\begin{figure}[t]
     \includegraphics[width=\linewidth]{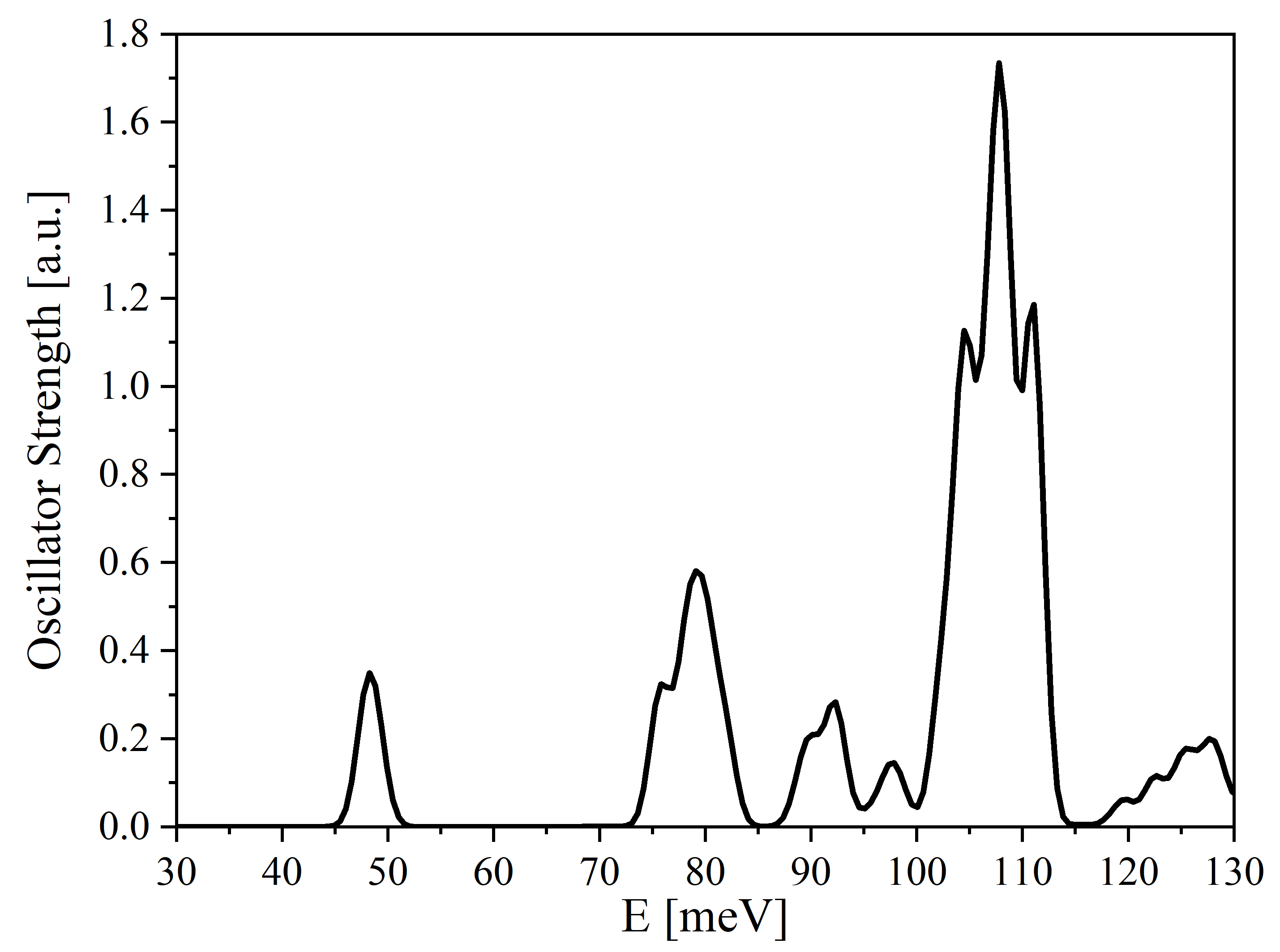}
    \caption{{Trion absorption spectrum.} Solid black line presents the Gaussian broadened absorption spectrum in the energy window from 30 to 130 meV.}
   \label{fig:Absorption}
\end{figure}

\subsection{Absorption}

\begin{figure}[t]
     \includegraphics[width=\linewidth]{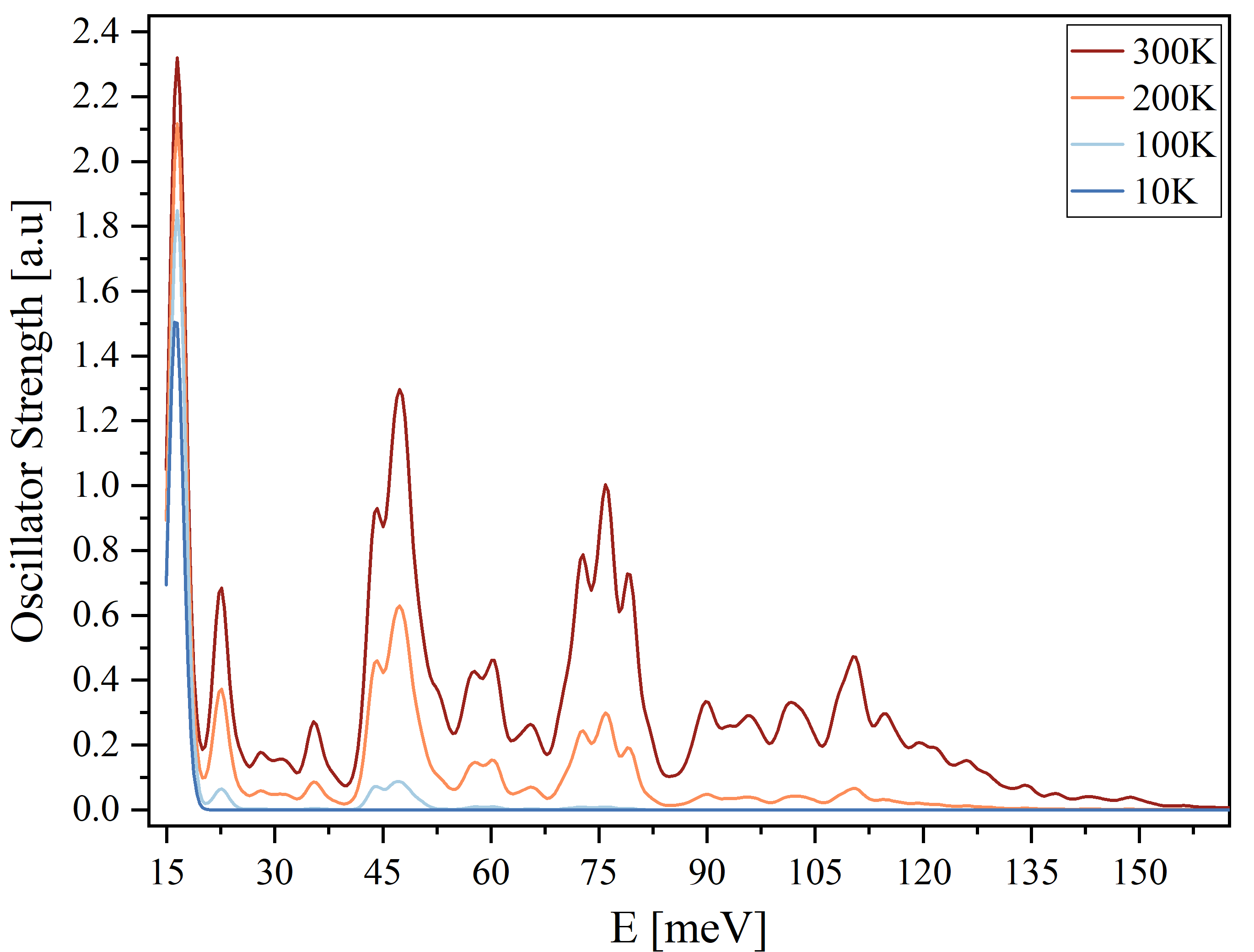}
    \caption{{Temperature-dependent emission spectrum.} The Gaussian broadened emission spectra presented for different temperatures. The temperature is given in Kelvin, Different colors mean different temperatures: blue - low temperatures, red - room temperature.}
   \label{fig:Emission}
\end{figure}

We now describe the absorption of light by a QD with an additional electron in the initial state and a negatively charged exciton in the final state. The probability of absorption of a photon is described by Fermi's golden rule, given 
by \cite{SaleemExciton2023, ozfidan_microscopic_2014}
\begin{equation}
    A\left(\omega\right) =\sum_{n,s}W_s\left|\braket{\psi_-^n|\hat{P}^\dagger|\psi^s}\right|^2 \delta\left(E^{n}-E^{s}-\omega\right), 
    \label{eq:FermiGoldenRuleAbsorption}
\end{equation}
where $\psi^s$ is the initial state. A photon of energy $\omega$ is absorbed, promoting an electron from the VB to the CB, and thus forming the trion complex in the state $\psi_-^n$. $W_s$ denotes the probability of the initial state $s$ being occupied at temperature $T$. The operator $\hat{P}^\dagger = \sum_{\alpha,i}D_{\alpha,i}c_{\alpha}^\dagger c_{i}$ is the polarization operator adding an electron-hole pair excitation while annihilating a photon weighted by the DME $D_{\alpha,i}$. We approximate the initial state $\psi^s$ as a single Slater determinant $\psi^s\approx\psi^{s_q}$ (fully occupied VB plus a single electron), where $q$ corresponds to the index of the QD state, as defined in Fig.~\ref{fig:SPQDEnergies}(a). Since absorption of a photon conserves $S_z$, we restrict these Slater determinants to have $S_z = \frac{1}{2}$. This can be written explicitly as:
\begin{equation}
    \psi^{s_q}=c_{q,\uparrow}^\dagger\ket{GS}.
    \label{eq:FermiGoldenRuleStates}
\end{equation}
At zero temperature there are only two states that are occupied by a single electron. These states correspond to $\psi^{s_1}=c_{1,\uparrow}^\dagger\ket{GS}$, and $\psi^{s_2}=c_{2,\uparrow}^\dagger\ket{GS}$, where $1$, $2$ are located at the bottom of the CB in the two non-equivalent valleys. The transition energy is calculated as 
$\omega = E^{n}-\epsilon_q-\Sigma_{q,q}$, where $\epsilon_q+\Sigma_{q,q}$ 
is the self-energy corrected single-particle energy of QD state $q$.

In Fig.~\ref{fig:Absorption} we present the absorption spectrum obtained from Eq.~(\ref{eq:FermiGoldenRuleAbsorption}). Similarly to the excitons in BLG QD \cite{SaleemExciton2023}, we find that the large absorption peak, corresponding to the optically dominant transitions $\gamma_b$ and $\gamma_c$ shown in Fig.~\ref{fig:DipoleMoments}, is observed at a higher energy. This transition is however broadened compared to the excitonic absorption peaks, since trion states are much more correlated, involve many more configurations than just excitons. We also note that the low-energy trion states are not completely dark, as the inclusion of trigonal warping, $\gamma_3$ and $\gamma_4$ in the bulk Hamiltonian, defined by Eq.~(\ref{eq:BulkH}), results in optically active low-energy trion states. This effect is similar to the brightening of the 1\textit{s} exciton in gated BLG \cite{henriques2022absorption}.

\subsection{Emission Spectrum}

We now move to a reverse process, in which the initial state is a trion state $\psi_-^n$. One of the two electrons in CB recombines with the hole, emitting a photon with energy $\omega$. The emission spectrum is described by Fermi's golden rule, given as \cite{cygorek2020atomistic,narvaez2005excitons}:
\begin{equation}
    A\left(\omega\right) =\sum_{n,s}W_n\left|\braket{\psi^s|\hat{P}|\psi_-^n}\right|^2 \delta\left(E^{s}+\omega-E^{n}\right), 
    \label{eq:FermiGoldenRuleEmission}
\end{equation}
where $W_n$ is the probability of a trion state $\psi_-^n$ being occupied at temperature $T$, defined as
\begin{equation}
    W_n = \frac{e^{\frac{-E_n}{k_BT}}}{\sum_me^{\frac{-E_m}{k_BT}}}.
    \label{eq:Thermaloccupation}
\end{equation}
We assume that the final state $\psi^s$ is given as a single Slater determinant defined by Eq.~(\ref{eq:FermiGoldenRuleStates}). Thus, the final state has an extra electron in any of the excited QD states. The transition energy is the same as in the absorption case. 

\begin{figure}[t]
     \includegraphics[width=\linewidth]{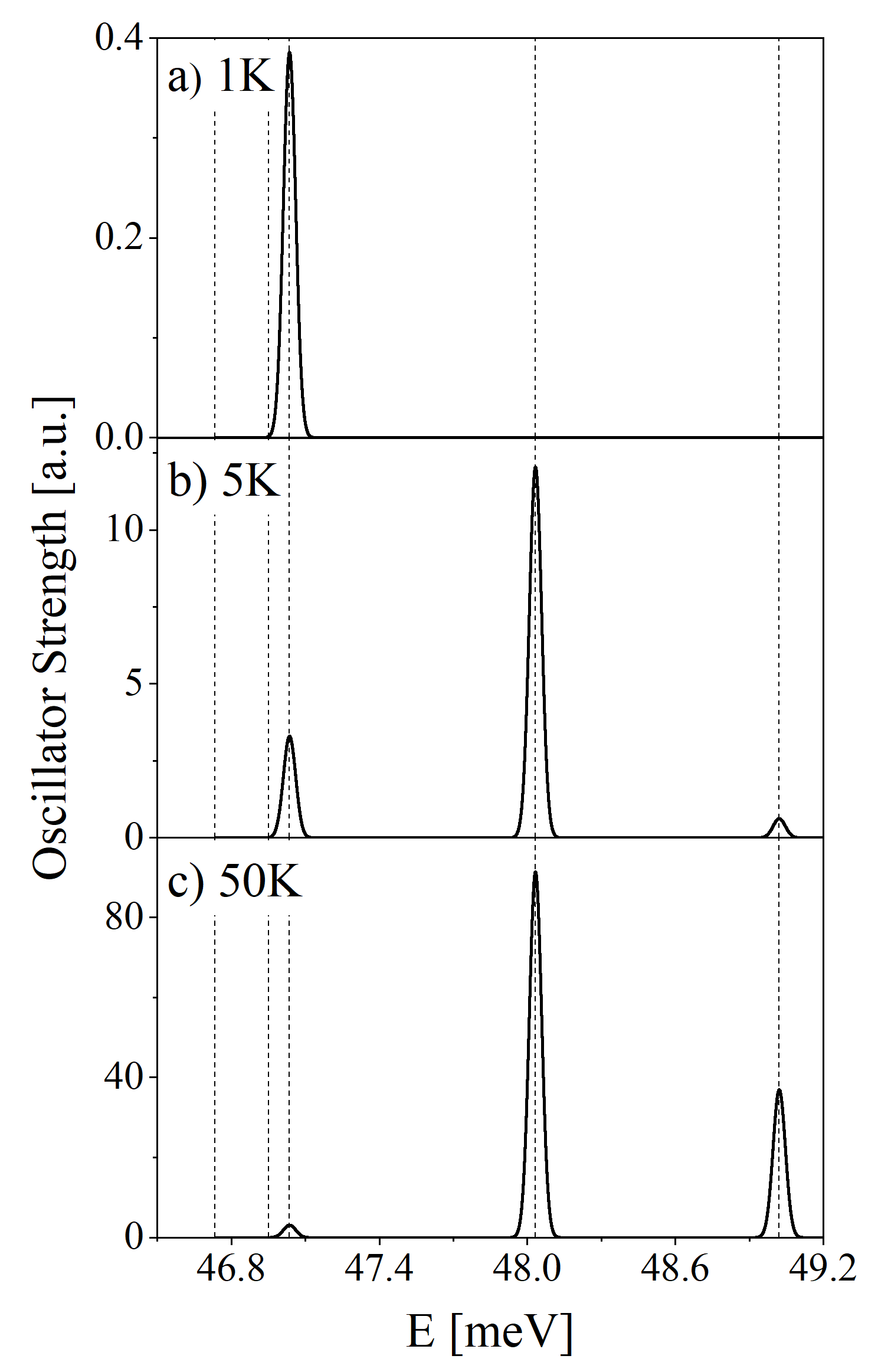}
    \caption{{} The emission spectrum from the 10 lowest-energy trion states presented for three different temperatures, (a) $T=1$~K, (b) $T=5$~K, (c) $T=50$~K, respectively. Solid lines denote the broadened emission spectrum, while the dashed vertical lines represent the possible low-energy recombination energies of the trion.}
   \label{fig:EmissionLowEnergy}
\end{figure}

Fig.~\ref{fig:Emission} shows the emission spectrum for different temperatures $T$. We find that in the emission spectrum the low-energy trions are bright rather than characterized by low intensity, unlike in the absorption picture. This effect occurs due to the large energy gap separating the low-energy manifold of 10 states from the higher energy trion states. For low temperatures, states at higher energies are not occupied. This results in the dominant emission peaks appearing at low energies. Furthermore, the energies of these bright low-energy trions can be tuned with the electric-field, making this QD a strong candidate for a single-photon emitter.

We now proceed to discussing the low-energy emission spectrum from the shell of $N=10$ low-energy trion states presented in Fig.~\ref{fig:TrionSpectrum}(b). Fig.~\ref{fig:EmissionLowEnergy} shows the low-energy emission spectrum for the temperatures of $1$~K, $5$~K, and $50$~K, respectively. The dashed lines represent the 5 doublets, corresponding to the initial trion states in the emission process. The first low-energy trion state appears to be dark in emission, for all temperatures. This can be understood by the fact that this state is characterized by the total spin $S=\frac{3}{2}$ and our final states defined in Eq.~(\ref{eq:FermiGoldenRuleStates}) have total spin $S=\frac{1}{2}$. Thus this emission process does not conserve total spin $S$, and as such is spin-forbidden. The second doublet in the manifold corresponds to a trion state in which the only possible recombination pathway involves an intervalley recombination. Since the intervalley DME are negligibly small, this state is also dark. The third doublet appears to be optically active and gives the largest emission peak when the temperature of $1$~K is considered. This state contains a pathway for an electron-hole pair to recombine within the same valley. The higher energy states as well contain an intravalley recombination path, however, they remain dark at $1$~K ($0.086$~meV). This energy is large enough to occupy the 3rd doublet, but the gap between the 3rd doublet and higher excited states is much larger than $0.1$~meV, leaving the 4th and 5th doublet thermally unoccupied. Once the temperature is increased to $5$~K, the emission occurs from the fourth doublet in the low-energy manifold, as presented in Fig.~\ref{fig:EmissionLowEnergy}(b). At the same time the emission peak from the third shell appears to be of low oscillator strength at the increased temperature. The fifth doublet does brighten at a higher temperature (Fig.~\ref{fig:EmissionLowEnergy}(c)), however it remains characterized by a low intensity compared to the emission peak from the fourth shell. The shifting of the emission peaks as a function of temperature would enable to partially resolve the fine structure of the trion in a BLG QD.



\section{Summary}

In summary, with the use of an \textit{ab initio}-based tight-binding approximation and the Bethe–Salpeter-like theory, we described negatively charged excitons confined in laterally gated BLG QDs. Unlike in conventional semiconducting QDs, trions in BLG QDs contain an electrically tunable fine structure of ten states arising from the valley and spin degrees of freedom. We predicted absorption to and emission from the trion states, where we identified valley dependent selection rules. We obtained bright low-energy trion states in emission and described how to extract the trion fine structure from the temperature dependence of the emission spectra. Interestingly, we found that the ground state trion is dark due to spin, contrary to the conventional self-assembled QDs \cite{Bayer2002TrionFineStructure,Witek2011TrionFineStructure,Wojs1995TrionQD}. Future work will include the effect of screening on Coulomb interactions in BLG QDs.



\section{Acknowledgments}
This work was supported by the Quantum Sensors Challenge Program QSP078 and Applied Quantum Computing AQC004 Challenge Programs at the National Research Council of Canada, NSERC Discovery Grant No. RGPIN- 2019-05714, and University of Ottawa Research Chair in Quantum Theory of Materials, Nanostructures, and Devices. We thank Digital Research Alliance Canada and Wrocław Centre for Networking and Supercomputing for computing resources. K.S acknowledges financial support from National Science Centre, Poland, under Grant No. 2021/43/D/ST3/01989. GB and YS acknowledge support by the Cluster of Excellence “Advanced Imaging of Matter” of the Deutsche Forschungsgemeinschaft (DFG) - EXC 2056-Project 390715994.



\section{Author Contributions}
K.S. and Y.S. contributed equally to this work.



\appendix*
\section{Trion Binding Energy}
\label{section:Appendix}

In this appendix we will calculate the trion binding energy. We define the binding energy of a negatively charged exciton as \cite{chafai2023insight}:
\begin{equation}
    E_b^T=E^T-E^{el}-E^X.
\label{eq:BindingEnergy}
\end{equation}
$E^T$ corresponds to the first bright trion state energy $E^T=40.6$~meV, as shown in Fig.~\ref{fig:TrionSpectrum}(b). $E^{el}$ describes the energy of a free electron corrected by the self-energy, which corresponds to the single-particle energy of an electron in the bottom CB state of the BLG QD renormalized by its self-energy. We determined this quantity to be equal to $-6.4$~meV. The value $E^T-E^{el}=47.03$~meV represents the energy of a photon emitted from the first bright trion state and corresponds to the third vertical dashed line (first emission peak) in Fig.~\ref{fig:EmissionLowEnergy}. The last component in Eq.~(\ref{eq:BindingEnergy}), $E^X$, denotes the energy of the first bright exciton state. It has been determined following the methodology described in our previous work \cite{SaleemExciton2023}, however, now with the inclusion of the non-zero matrix elements $\gamma_3$ and $\gamma_4$ in the bulk Hamiltonian described in Eq.~(\ref{eq:BulkH}) and with the screening constant $\kappa=3.9$. We found this value to be $E^X=53.2$~meV. This analysis allowed us to determine the trion binding energy as $E_b^T=-6.4$~meV. We obtained a bound trion, whose emission maximum is located at a lower energy than the exciton peak.


\newpage*

%

\end{document}